\documentclass{aastex63}
\usepackage[T1]{fontenc}
\pdfoutput=1
\usepackage[latin9]{inputenc}
\usepackage{bm}
\usepackage{amstext}
\usepackage{graphicx}

\makeatletter
\usepackage{savesym}
\savesymbol{iint}
\savesymbol{iiint}
\savesymbol{iiiint}
\usepackage{amsmath}
\usepackage{array}
\usepackage{multirow}
\usepackage[caption=false]{subfig}

\makeatother

\usepackage{babel}
\begin{document}

\global\long\def\grad{\bm{\nabla}}%
\global\long\def\curl{\bm{\nabla}\times}%

\title{A linear model for inertial modes in a differentially rotating Sun}
\author[0000-0001-6433-6038]{Jishnu Bhattacharya}
\affiliation{Center for Space Science, New York University Abu Dhabi, Abu Dhabi, P.O. Box 129188, UAE}
\author[0000-0003-2536-9421]{Chris S. Hanson}
\affiliation{Center for Space Science, New York University Abu Dhabi, Abu Dhabi, P.O. Box 129188, UAE}
\author[0000-0003-2896-1471]{Shravan M. Hanasoge}
\affiliation{Department of Astronomy and Astrophysics, Tata Institute of Fundamental Research, Mumbai - 400005, India}
\affiliation{Center for Space Science, New York University Abu Dhabi, Abu Dhabi, P.O. Box 129188, UAE}
\author[0000-0002-3943-6827]{Katepalli R. Sreenivasan}
\affiliation{Center for Space Science, New York University Abu Dhabi, Abu Dhabi, P.O. Box 129188, UAE}
\affiliation{New York University, NY, USA 10012}

\begin{abstract}
    Inertial wave modes in the Sun are of interest owing to their potential to reveal new insight into the solar interior. These predominantly retrograde-propagating modes in the solar subsurface appear to deviate from the thin-shell Rossby-Haurwitz model at high azimuthal orders. We present new measurements of sectoral equatorial inertial modes at $m>15$ where the modes appear to become progressively less retrograde compared to the canonical Rossby-Haurwitz dispersion relation in a co-rotating frame. We use a spectral eigenvalue solver to compute the spectrum of solar inertial modes in the presence of differential rotation. Focussing specifically on equatorial Rossby modes, we find that the numerically obtained mode frequencies lie along distinct ridges, one of which lies strikingly close to the observed mode frequencies in the Sun. We also find that the $n=0$ ridge is deflected strongly in the retrograde direction. This suggests that the solar measurements may not correspond to the fundamental $n=0$ Rossby-Haurwitz solutions as was initially suspected, but to a those for a higher $n$. The numerically obtained eigenfunctions also appear to sit deep within the convection zone --- unlike those for the $n=0$ modes --- which differs substantially from solar measurements and complicates inference.
\end{abstract}
\section{Introduction}

Global-scale oscillations such as Rossby waves \citep{Rossby1945} are characteristic of rotating atmospheres, and they have been studied in terrestrial settings \citep{pedlosky1987,pedlosky2003}, as well as in astrophysical settings \citep{Lou2000,Lanza2009, Zaqarashvili2021}. Waves with similar temporal and spatial characteristics have also recently been detected on the Sun \citep{Loeptien2018, Liang2019, Hanasoge2019, Mandal2020, Hanson2020, Proxauf2020, Mandal2021, Waidele2023}. A key characteristic of these largely toroidal oscillation modes is that the primary restoring contribution to sustain them is provided by the Coriolis force. Several authors have expounded on the linearized theory of inertial modes in slowly rotating stars \citep{Provost1981, Saio1982}. Typically, the angular profiles of modes on the Sun are interpreted in a basis of spherical harmonics, $Y_{\ell,m}(\theta,\phi)$, where $\ell$ is the angular degree and $m$ is the azimuthal order, and $\theta$ and $\phi$ represent the co-latitude and azimuth respectively. The dominant modes that are observed on the Sun are sectoral, that is, most of the power in the mode is concentrated in the channel $\ell=m$, and these modes are symmetric about the equator. More recently, observations of equatorially antisymmetric modes of oscillation by \citet{Hanson2022} have been interpreted as the fundamental full-sphere oscillations in the solar convection zone corresponding to $\ell=m+1$ \citep{Triana2022}. Several early studies had chosen to analyze linear oscillations about a uniformly rotating background, but the differential rotation of the Sun may be expected to contribute significantly to the spectrum of modes. This is because, firstly, the Doppler shift encountered by a mode with an azimuthal order $m$ --- given by $m\Delta\Omega$ for a constant rate of differential rotation $\Delta\Omega$ about a reference rotation rate of $\Omega_0$ --- may dominate the canonical dispersion relation of sectoral Rossby modes $-2\Omega_0/(m+1)$ at high $m$, and secondly, non-uniform rotation couples modes across the harmonic orders to a much larger degree than uniform rotation, which makes mode identification and labelling challenging. The former leads to critical latitudes, where the inviscid eigenvalue problem becomes formally singular \citep{Watts2004, Gizon2020}. In certain conditions, differential rotation might induce instabilities in the modes of oscillation \citep{Zaqarashvili2021,Fournier2022}. The use of these modes in helioseismology, therefore, requires a comprehensive understanding of the dispersion relation and stability of inertial modes in a solar-like background rotation.

Several authors \citep{Gizon2020, Fournier2022, Bekki2022a, Bekki2022b} have carried out numerical analysis of inertial modes in the presence of a solar-like differential rotation profile. Each set of authors reported r-mode dispersion relations where the real parts of the frequencies were close to the central frequencies of observed solar inertial modes. \citet{Gizon2020} suggest that equatorial modes are confined between critical latitudes, and that other categories of modes --- such as critical-latitude modes and high-latitude modes --- might exist in a differentially rotating medium. These modes have putatively been detected on the Sun as well \citep{Gizon2021}. \citet{Hathaway2013}, and later \citet{Bogart2015} and \citet{Hathaway2021}, found evidence for polar vortices, which might correspond to a self-excited, high-latitude $m=1$ mode driven by baroclinicity \citep{Bekki2022b}. \citet{Bekki2022a} have detected several classes of inertial modes in nonlinear simulations of rotating convection, including equatorial modes, the high-latitude $m=1$ feature, as well as columnar convective modes. Of these, equatorial inertial modes are the easiest to identify, as their frequencies are close to the canonical thin-shell Rossby-Haurwitz dispersion relation for sectoral modes ($\omega=-2\Omega_0/(m+1)$ for a uniform rotation rate $\Omega_0$).

In this work, we compute the spectrum of the equatorial inertial modes numerically in the anelastic approximation by including a solar-like differential rotation profile. This approach differs from previous studies, where differential rotation had been studied either as a radius-dependent latitudinal profile \citep{Fournier2022}, or as a radially and latitudinally varying profile that is confined to a truncated radial domain \citep{Bekki2022b}. The latter of these studies comes the closest to our analysis, although the specifics of the analysis differ in the way compressibility is addressed, the profile of rotation that is used in the analysis, the tracking rate used to compute the mode frequencies, as well as in the numerical approach (finite difference in a meridional plane in \citep{Bekki2022b}, whereas spectral in ours). We use a Petrov-Galerkin discretization of the equations of mass, momentum and energy, following the approach presented by \citet{Bhattacharya2023}. A spectral approach often leads to exponential convergence for smooth functions \citep{Gottlieb1978}, which may reduce the resolutions required to obtain convergent solutions, and consequently make the eigenvalue problem in a non-uniform background more tractable. The nominally higher accuracy expected from spectral methods also affords more trust in the solutions.  One of the main questions that we seek to answer in this work is why the dispersion relation of the Rossby waves that are observed on the Sun is consistent with that seen in a thin-shell such as the earth's atmosphere, when the radial extent of the solar convection zone spans many scale heights, and is therefore far from a thin shell. Furthermore, the measured dispersion relation appears to progressively deviate from the Rossby-Haurwitz model (becoming less retrograde than expected at $m>10$), and we explore whether this may be explained in terms of the solar differential rotation.

\section{Wave equation}

In this analysis, we use a spherical coordinate system, with $r$ representing the radius, and $\theta$ and $\phi$ standing for the co-latitude and azimuth respectively. We model waves as small perturbations in a spherical shell with a lower radial bound denoted by $r_\text{in}$ and an outer bound $r_\text{out}$. We choose the background to correspond to Model S \citep{ChristensenDalsgaard1996} that is augmented by a super-adiabaticity profile inspired by \citet{Rempel2005} (see \citep{Bhattacharya2023} for details). The background profile may be represented in terms of its radially varying profiles of density $\bar{\rho}(r)$, pressure $\bar{p}(r)$, temperature $\bar{T}(r)$, gravitational acceleration $\mathbf{g}(r)$ that is directed inwards, as well as a radially and latitudinally varying profile of entropy $S(r,\theta)$. We denote perturbations to the background parameters by using primes as superscripts. We denote the wave velocity by $\mathbf{u}(r,\theta,\phi,t)$, and its vorticity $\curl\mathbf{u}(r,\theta,\phi,t)$ by $\bm{\omega}(r,\theta,\phi,t)$. Wherever unambiguous, we drop the spatial and temporal coordinates in the interest of brevity.

We assume that the medium is close to the anelastic limit \citep{ Gough1969, Gilman1981, Braginsky1995}, where the equation of mass continuity takes the form
\begin{align}
\grad\cdot\left(\bar{\rho}\mathbf{u}\right)=0.
\end{align}
In this scenario, we assume that  fluid velocities are highly subsonic, and therefore filter sound waves out of the analysis. This decouples pressure and density from the fluid velocity. Pressure perturbations do not evolve in time and may be solved for at each instant given the velocity profile, while the density perturbation may be obtained from an equation of state. This reduces the number of unknowns that we solve for to the velocity profile and entropy perturbation $S^\prime$. Imposition of the anelastic approximation further restricts the number of independent components of the velocity field to two.

We represent the rotation velocity of the medium by $\bm{\Omega}=\Omega(r,\theta)\,\mathbf{e}_z$, and assume that this is being tracked in a frame rotating at a constant angular velocity $\mathbf{\Omega}_0=\Omega_0\,\mathbf{e}_z$. In a uniformly rotating medium where $\Omega$ is a constant in space, one may choose the tracking rate $\Omega_0$ to be equal to $\Omega$, and analyze wave propagation in a co-rotating frame. In the Sun, however, differential rotation contributes additional terms to the wave equation. We denote the $\mathbf{e}_z$-component of the differential rotation velocity $\Delta\bm{\Omega}=\bm{\Omega}-\mathbf{\Omega}_0$ by $\Delta\Omega(r,\theta)$.

We represent the momentum equation in terms of the velocity $\mathbf{u}$, the vorticity $\bm{\omega}=\curl\mathbf{u}$, and the entropy perturbation $S^\prime$, in the form presented by \citet{Lantz1992} and \citet{Braginsky1995}. The Navier-Stokes equation in a rotating frame in the anelastic approximation may be expressed as
\begin{align}
    \partial_{t}\mathbf{u}=\mathbf{u}\times\bm{\omega}-2\boldsymbol{\Omega}\times\mathbf{u}-\grad\left(\frac{p^{\prime}}{\bar{\rho}}+\frac{1}{2}\mathbf{u}^{2}\right)-\frac{S^{\prime}}{c_{p}}\mathbf{g}+\frac{1}{\bar{\rho}}\mathbf{F}^{\nu},
    \label{eq:momentum}
\end{align}
where the viscous force term $\mathbf{F}^{\nu}$ is given by 
\begin{align}
    \mathbf{F}^{\nu} = \nu\left[\grad\cdot\left(\bar{\rho}\left(\grad\mathbf{u}+\left(\grad\mathbf{u}\right)^{T}\right)\right)-\frac{2}{3}\grad\left(\bar{\rho}\grad\cdot\mathbf{u}\right)\right],
\end{align}
and we assume that $\nu$ is a constant. We have applied the vector identity $\mathbf{u}\cdot\grad\mathbf{u}=\frac{1}{2}\grad\left|\mathbf{u}\right|^{2}-\mathbf{u}\times\boldsymbol{\omega}$ to obtain terms on the right-hand side. This form of the equation emphasizes that only the $\mathbf{u}\times\boldsymbol{\omega}$ term of the two contributes to the time-evolution of vorticity, as the curl of a gradient is zero.

\citet{Bhattacharya2023} had analyzed the terms that correspond to a uniformly rotating background, and in this work, we focus on the additional terms arising from differential rotation. The $\mathbf{u}\times\bm{\omega}$ term is second-order in $\mathbf{u}$ in a uniformly rotating medium. In a differentially rotating medium, however, there arises a first-order contribution due to a coupling between the background rotation and the fluid velocity.
We may expand the velocity field in terms of the intrinsic fluid velocity $\mathbf{u}_f$ and the velocity $\mathbf{u}_\Omega$ that is associated with differential rotation as
$\mathbf{u}= \mathbf{u}_{f}+\mathbf{u}_\Omega$, where $\mathbf{u}_\Omega=\Delta\bm{\Omega}\times\mathbf{r}$. Defining $\boldsymbol{\omega}_{\Omega}=\curl\mathbf{u}_\Omega$, we may expand the advection term as
\begin{align}
    \mathbf{u}\times\boldsymbol{\omega}&\approx\mathbf{u}_{f}\times\boldsymbol{\omega}_{\Omega}+\mathbf{u}_{\Omega}\times\boldsymbol{\omega}_{f},
\end{align}
where we have assumed that the centrifugal component is negligible in comparison with gravity, and we have ignored the second-order terms in $\mathbf{u}_f$ as well. In addition to the advection term, the Coriolis force due to differential rotation may be expressed as $-2\Delta\boldsymbol{\Omega}\times\mathbf{u}_{f}$. Collectively, the net contribution of differential rotation becomes
\begin{align}
 -\left(\boldsymbol{\omega}_{\Omega}+2\Delta\boldsymbol{\Omega}\right)\times\mathbf{u}_{f}+\mathbf{u}_{\Omega}\times\boldsymbol{\omega}_{f}.
 \label{eq:ucrossomegaterms}
\end{align}
The second term in Equation \eqref{eq:ucrossomegaterms} contributes $-im\Delta\Omega\,\omega_{fr}$ to the time-evolution of the radial component of vorticity, which we may recognize as a Doppler shift. Since this term is linear in $m$ whereas the sectoral Rossby-Haurwitz dispersion relation $2\Omega/(m+1)$ varies inversely with $m$, we may expect the Doppler-shift to contribute significantly for $m\gg1$. To analyze the impact of the first term in Equation \eqref{eq:ucrossomegaterms}, we note that if $\Delta\Omega$ is a constant in space (which we may denote by $\Delta\Omega_\mathrm{const}$), the corresponding vorticity may be expressed as $\bm{\omega}_\Omega=2\Delta\bm{\Omega}_\mathrm{const}$. The first term in Equation \eqref{eq:ucrossomegaterms} consequently becomes $-4\Delta\bm{\Omega}_\mathrm{const}\times\mathbf{u}_{f}$, which acts as an augmented Coriolis force. The impact of this term on the spectrum would be a scaling of the rotation velocity in the Rossby-Haurwitz dispersion relation. The rotation profile in the Sun is far from constant, but we may interpret the result in terms of an appropriately averaged rotation rate over the domain.

We do not consider the contribution of differential rotation to the viscous force. \citet{Fournier2022} have demonstrated that the real and the imaginary parts of the eigenfrequencies are related, and increasing the Ekman number induces a change in both (especially for Ekman numbers above $10^{-3}$), although, out of all the solutions, the r mode is the least affected by an increase in the Ekman number. \citet{Bekki2022b} also find a mild dependence of the equatorial Rossby modes on the viscosity coefficient. In this analysis, we are in a weak-viscosity regime ($\nu=5\times10^{11}\,\text{cm}^2/\text{s}$, corresponding to an Ekman number of around $10^{-5}$), and therefore this impact may be expected to be insignificant. In any case, our assumption of the model of how turbulent viscosity impacts inertial waves is simplistic, and more realistic models that incorporate the depth-dependence of viscosity might be necessary to conclusively comment on the impact of differential rotation on the line widths on modes (see e.g., \citet{MunozJaramillo2011} for a discussion on various models of depth-dependent diffusivity). However, it remains to be determined whether these sorts of subtleties induce observable changes to the modes. 

The energy equation may be expressed in terms of the entropy perturbation $S^\prime$ and the wave velocity $\mathbf{u}_f$ as
\begin{align}
    \partial_{t}S^{\prime}=-\mathbf{u}_\Omega \cdot \grad S^{\prime}-\mathbf{u}_{f}\cdot\grad S+\kappa\frac{1}{\bar{\rho}\bar{T}}\grad\cdot\left(\bar{\rho}\bar{T}\grad S^{\prime}\right).
    \label{eq:entropy}
\end{align}
We have disregarded other terms such as radiative and viscous heating in this analysis, but such terms may be included in a straightforward manner. The radiative heating term contributes to thermal transport, so an improved treatment of the energy equation would include such a term \citep{Featherstone2016b}. We may simplify Equation \eqref{eq:entropy} further by noting that the radial derivative of the entropy profile may be expressed in terms of the super-adiabatic gradient $\delta$ as
$\partial_{r}S=-\gamma\delta/H_{\rho}$,
where $\gamma$ is the adiabatic exponent, and $H_{\rho}$ is the density scale-height. An estimate of the latitudinal gradient of entropy is found to be $\partial_\theta S=2(c_{p}/g)\, r^{2}\sin\theta\,\bm{\Omega}_{0}\cdot\grad\Delta\Omega$ by requiring thermal-wind balance. \citet{Bekki2022b} find that this term contributes significantly to the instability of the $m=1$ high-latitude mode, and, as a consequence, the degree of latitudinal non-adiabaticity may be probed through solar inertial modes. The influence of this term on equatorial modes may be limited \citep{Bhattacharya2023}, since the term is proportional to $\partial_\theta \Delta\Omega$, which is small near the equator compared to the reference rotation rate $\Omega_0$, and increases with latitude (see Fig \ref{fig:rotvel}). Owing to computational limitations, we do not include the latitudinal entropy gradient term in our analysis.

We choose a poloidal-toroidal decomposition of the wave velocity $\mathbf{u}_f$ that automatically satisfies the continuity equation. The azimuthal symmetry of the rotational angular velocity $\bm{\Omega}$ implies that $\rho\mathbf{u}_\Omega$ is divergence-free, so the continuity equation translates to that for the wave velocity $\grad\cdot\left(\rho\mathbf{u}_f\right)=0$. We express the wave velocity $\mathbf{u}_f$ in terms of stream functions as
\begin{align}
    \bar{\rho}\mathbf{u}_f=\grad\times\grad\times\left(\bar{\rho}\,W\left(\mathbf{x}\right)\hat{r}\right)+\grad\times\left(\bar{\rho}\,V\left(\mathbf{x}\right)\hat{r}\right),
    \label{eq:VWexpansion}
\end{align}
where $V(\mathbf{x})$ corresponds to the radial component of the wave vorticity, and $W(\mathbf{x})$ corresponds to the radial component of the wave velocity. We may express the velocity components in terms of the stream functions as
\begin{equation}
\begin{aligned}v_{r} & =-\frac{1}{r^{2}}\nabla_{h}^{2}W\\
v_{\theta} & =\frac{1}{r}\left(\frac{1}{\bar{\rho}}\partial_{\theta}\partial_{r}\left(\bar{\rho}W\right)+\frac{1}{\sin\theta}\partial_{\phi}V\right),\\
v_{\phi} & =\frac{1}{r}\left(\frac{1}{\bar{\rho}}\frac{1}{\sin\theta}\partial_{\phi}\partial_{r}\left(\bar{\rho}W\right)-\partial_{\theta}V\right),
\end{aligned}
\end{equation}
where $\nabla_{h}$ represents the angular component of the Laplacian operator. This suggests that in the special case where the field is purely toroidal, sectoral, and featuring a single azimuthal order $m$, the stream function $V$ would have an angular profile proportional to $\sin^m\theta$, and $v_\theta$ would have a latitudinal dependence given by $\sin^{m-1}\theta$. In this case, $v_\theta$ would be symmetric about the equator, whereas $v_\phi$ would be antisymmetric.

Substituting Equation \eqref{eq:VWexpansion} in \eqref{eq:momentum} and \eqref{eq:entropy}, we obtain coupled equations for $V(\mathbf{x})$, $W(\mathbf{x})$ and $S^\prime(\mathbf{x})$. Transforming to temporal frequency, the equations reduce to an eigenvalue problem where the eigenvalues are mode frequencies $\omega$, and the eigenvectors are $[V(\mathbf{x}),W(\mathbf{x}),S^\prime(\mathbf{x})]$.

\section{Numerical Technique}

\subsection{Discretizing operators}

\citet{Bhattacharya2023} had presented an approach to discretize the eigenvalue equation for inertial waves in a spectral basis spanned by Chebyshev polynomials in the radial direction, and spherical harmonics in the lateral coordinates. They applied the approach to a background medium that was either rotating uniformly or differentially in radius, but latitudinally uniform. In this work, we extend their approach to a latitudinally differentially rotating background, bearing in mind that the solar rotation profile is azimuthally symmetric.
We choose to work in spherical polar coordinates, with the North Pole directed  along the Sun's axis of rotation. In this convention, $\theta$ represents the co-latitude, and $\phi$ the azimuth. We limit our radial domain to $r_\mathrm{in}=0.6R_\odot$ to $r_\mathrm{out}=0.985R_\odot$ owing to concerns of numerical expense. We define the non-dimensional radius $\hat{r}$ as
\begin{align}
r_{\text{mid}}&=\frac{r_{\text{out}}+r_{\text{in}}}{2},\\
\hat{r}&=\frac{r-r_{\text{mid}}}{r_{\text{out}}-r_{\text{mid}}},    
\end{align}
which maps the radial domain $[r_\mathrm{in},r_\mathrm{out}]$ to $[-1,1]$.

\begin{figure}
    \centering
    \includegraphics[scale=0.8]{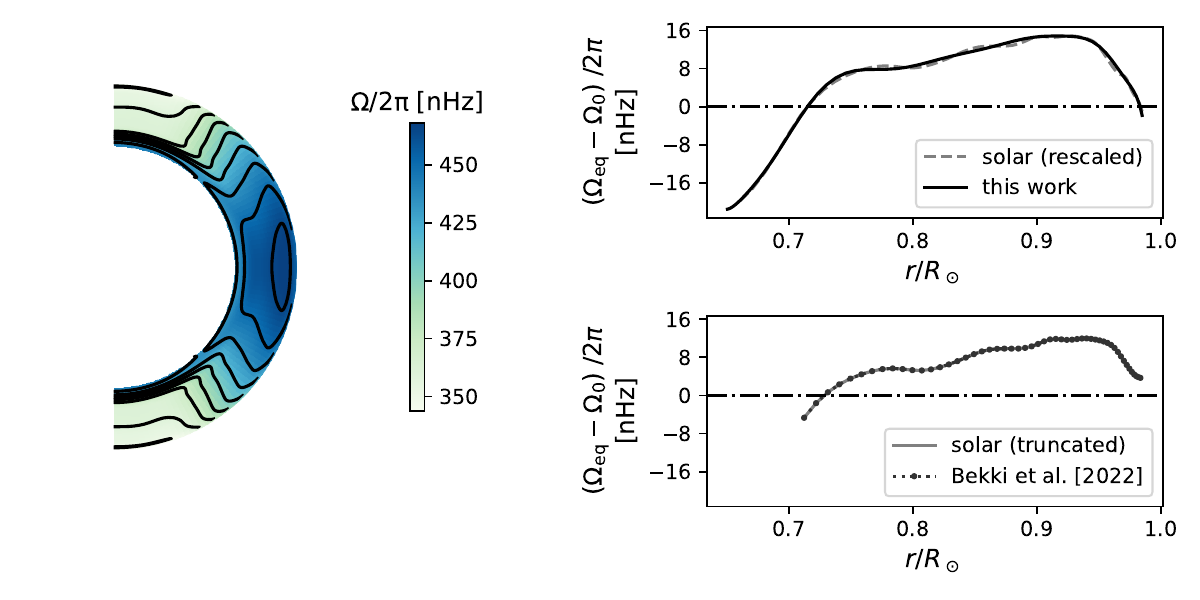}
    \caption{Left: Profile of the smoothed solar rotation velocity $\Omega(r,\theta)$ used in the analysis. Top right: Equatorial cross-section of the differential rotation $(\Omega(r,\theta=\pi/2) - \Omega_0)/2\pi$. The smoothed profile (black) is used in the analysis. For reference, the dashed gray line depicts the rescaled solar rotation profile. Bottom right: the rotation profile used by \citep{Bekki2022b}, and the corresponding solar rotation profile truncated to the radial domain $[0.71R_\odot,0.985R_\odot]$ that is used by the authors. The reference rate $\Omega_0$ in the bottom right panel is chosen to be the Carrington rate $2\pi\times456\,\text{nHz}$ to match that in the work.}
    \label{fig:rotvel}
\end{figure}

The Doppler shift induced by differential rotation on the inertial-wave spectrum depends critically on the rotation profile, and we take extra care to incorporate the profile in its entirety in our domain. We choose a second scaled radial coordinate
\begin{align}
r_{\text{rot}} & =r_{\text{in}}+\frac{r-r_{\text{in}}}{r_{\text{out}}-r_{\text{in}}}\left(R_{\odot}-r_{\text{in}}\right)
\end{align}
which maps the domain $[r_\mathrm{in},r_\mathrm{out}]$ to $[r_\mathrm{in},R_\odot]$. In terms of this radius, we define our modified rotation profile $\tilde{\Omega}\left(r,\theta\right)$ as
\begin{align}
    \tilde{\Omega}\left(r,\theta\right)=\Omega\left(r_{\text{rot}},\theta\right),
\end{align}
which squeezes the outer envelope of the rotation profile in the Sun \citep{Larson2018} to the radial domain used in this analysis. This procedure ensures that the impact of the near-surface shear layer is incorporated in the analysis while avoiding the steep density gradient associated with $r>0.985R_\odot$. From this point onward, we treat the modified profile $\tilde{\Omega}\left(r,\theta\right)$ as the one that features in Equation \eqref{eq:momentum}. We may expand this profile in a Chebyshev-Legendre tensor-product basis as
\begin{align}
    \tilde{\Omega}(r,\theta)&=\sum_{n=0}^\infty\sum_{\ell=0}^\infty \Omega_{n\ell} T_n(\hat{r}) \hat{P}_\ell\left(\cos\theta\right),
    \label{eq:rotationexpansion}
\end{align}
where $T_n(\hat{r})$ is the Chebyshev polynomial of degree $n$, and $P_\ell\left(\cos\theta\right)$ is the normalized Legendre polynomial of degree $\ell$. The $\ell=0$ term in the sum corresponds to the latitudinally uniform component of rotation, that had been studied by \citet{Bhattacharya2023}, of which the $n=0$ term corresponds to uniform rotation. We compute the expansion coefficients $\Omega_{n\ell}$ by firstly fitting a smoothing spline to the rotation angular velocity, and secondly performing an adaptive orthogonal polynomial fit to the spline. The smoothing procedure reduces the number of coefficients that contribute in Equation \eqref{eq:rotationexpansion}. We set our reference tracking rate to $\Omega_0 = 2\pi \times 453.1\,\text{nHz}$ to match the commonly used mean equatorial rotation rate at the solar surface that has been determined using f-mode seismology \citep{Schou1999}. This matches the rate used by \citet{Loeptien2018} and subsequent papers. We plot the rotation profile, as well as the equatorial profile of the differential component of the profile, in Figure \ref{fig:rotvel}.

We expand the stream functions $V$ and $W$ and the entropy $S^\prime$ in a Chebyshev-spherical-harmonic basis as
\begin{align}
\begin{aligned}V\left(\mathbf{x}\right) & =\sum_{\ell m}V_{n\ell m}T_{n}\left(\hat{r}\right)\hat{P}_{\ell m}\left(\cos\theta\right)\exp\left(im\phi\right),\\
W\left(\mathbf{x}\right) & =\sum_{\ell m}W_{n\ell m}T_{n}\left(\hat{r}\right)\hat{P}_{\ell m}\left(\cos\theta\right)\exp\left(im\phi\right),\\
S^{\prime}\left(\mathbf{x}\right) & =\sum_{\ell m}S_{n\ell m}^{\prime}T_{n}\left(\hat{r}\right)\hat{P}_{\ell m}\left(\cos\theta\right)\exp\left(im\phi\right),
\end{aligned}
\label{eq:VWSchebyshbasis}
\end{align}
where $T_{n}\left(\hat{r}\right)$ are Chebyshev polynomials and $\hat{P}_{\ell m}\left(\cos\theta\right)$ are normalized associated Legendre polynomials. Substituting Equation \eqref{eq:VWSchebyshbasis} into  \eqref{eq:momentum} and \eqref{eq:entropy}, we recast the eigenvalue problem in terms of the expansion coefficients $V_{n \ell m}$, $W_{n \ell m}$ and $S^\prime_{n \ell m}$. Azimuthal symmetry of the background leads to the equations decoupling in the azimuthal order $m$, and we may solve a set of smaller eigenvalue problems in each $m$, where the eigenvectors are the collated coefficients $[V_{n\ell m},W_{n\ell m},S^\prime_{n\ell m}]$ for that $m$.

We describe the approach to computing the matrix elements of the angular terms in Appendix \ref{app:matrixelem}. The discretized representation of a separable operator is given by the tensor product of the corresponding radial and angular matrices, whereas for a non-separable operator such as the rotation profile, we may express the result as a sum over separable terms, as in Equation \eqref{eq:rotationexpansion}. We use the Julia package ApproxFun.jl \citep[version 0.13,][]{Olver2014} to obtain this discretized representation, and subsequently use the LAPACK wrappers presented by the Julia standard library LinearAlgebra to perform the eigen-decomposition (using Julia version 1.9). The code for this is published freely on GitHub under the MIT license \footnote{https://github.com/jishnub/RossbyWaveSpectrum.jl}.

\subsection{Boundary conditions}

We impose impenetrable, stress-free boundary conditions at the radial extremities. We also require that there be no entropy flux across the boundaries. This may be expressed in terms of the velocity components as
\begin{align}
    u_{r} = \partial_{r}\left(\frac{u_{\theta}}{r}\right)=\partial_{r}\left(\frac{u_{\phi}}{r}\right)=\partial_r S^\prime=0,\quad\text{on}\;r=r_{i},\quad\text{and}\;r=r_{o}.
\end{align}
We enforce the boundary conditions by following a basis-recombination approach as presented by \citet{Bhattacharya2023}.

We do not impose additional boundary conditions at the poles aside from the ones that naturally emerge from the choice of spherical harmonics as the basis, which ensures smoothness of the stream functions at the poles. This may be seen as a behavioral boundary condition rather than a numerical one \citep{Boyd2000}.

\section{Data Analysis}
The mode properties (frequency, line width etc.) of the sectoral Rossby waves have been studied extensively up to $m=15$ \citep[][and subsequent papers]{Loeptien2018}. Here, we aim to characterize higher-order modes, i.e., $m>15$. There are a number of factors that have made it difficult to analyze these higher-order modes. Firstly, the amplitude of the modes decreases with increasing $m$, leading to poor signal-to-noise ratios for the fitting of the modes. This makes it difficult to characterize these high degree modes with techniques that have poor signal-to-noise across the spectrum (e.g. time-distance measurements). Furthermore, some data products have a low spatial Nyquist frequency. For instance, the GONG ring diagram pipeline \citep{Corbard2003} has tiles separated by $15^\circ$, leading to a spatial Nyquist of $m=12$. Beyond $m=15$ aliasing pollutes the spectrum and makes mode characterization impossible for the Global Oscillation Network Group (GONG) data. However, the flow maps from the Helioseismic and Magnetic Imager (HMI) ring pipeline have a Nyquist frequency of $m=24$, and here we average all depths down to $\sim20\,\text{Mm}$ below the surface to improve the signal-to-noise ratio \citep[e.g.][]{Hanson2022}. Global mode coupling and local correlation tracking (LCT) are not as limited in the spatial frequencies, though as reported by \citet{Loeptien2018} the latter has weak power at high $m$. Here, we utilize toroidal flows computed from the global mode coupling analysis (MCA) for both odd and even $m$ \citep{Mandal2021}, as well as the HMI ring diagram analysis (RDA) flows averaged for all depths down to $\sim20$\,\text{Mm}. 

\begin{figure}
    \centering
    \includegraphics[width=0.7\linewidth]{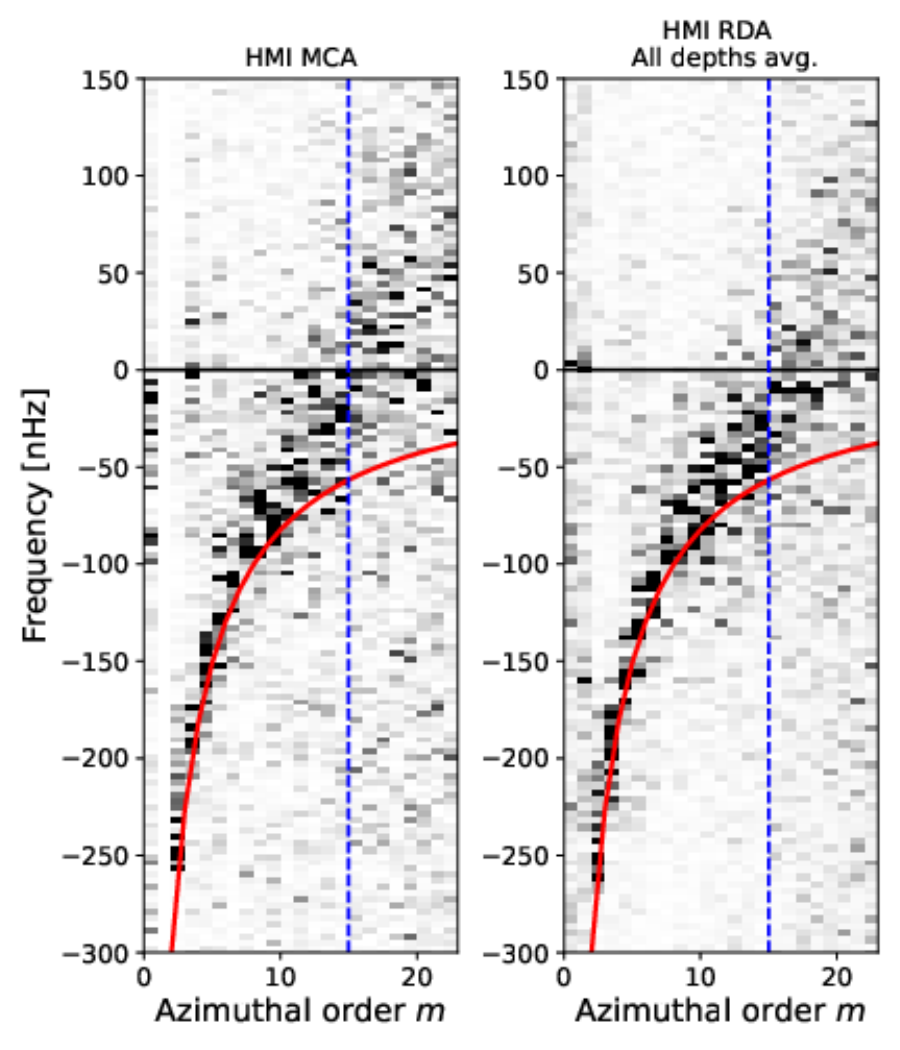}
    \caption{Sectoral power spectrum ($\ell=m$) of the radial vorticity, two different helioseismic techniques. Left panel: Radial vorticity spectrum computed using the global Mode Coupling Analysis (MCA), with even azimuthal orders \citep{Mandal2021}, on HMI data (2010-2020). Right panel: Radial vorticity spectrum computed through ring diagram analysis and averaging all depths, e.g. \citet{Hanson2022}. The red line represents the sectoral Rossby dispersion in the thin shell approximation ($-2\Omega_0/(m+1)$), while the blue dashed line shows the maximum azimuthal order to which previous studies have characterized the modes.}
    \label{fig:spectra}
\end{figure}

Figure~\ref{fig:spectra} shows the radial vorticity (toroidal flow) power spectrum of the HMI MCA and RDA. In both cases, the mode ridge begins to shift away from the thin-shell dispersion relation, becoming more prograde with increasing $m$. At these relatively high azimuthal order, any small Doppler shift will become significant due to the factor $m$ in the frequency shift. In both the MCA and RDA, there is an apparent mode power beyond $m=15$ up to $m=21$, where the modes appear to become slightly prograde (relative to the equatorial rotation frame of reference). Beyond the azimuthal order $m=21$, there is no discernible signal in any of the measurements. 

\begin{figure}
    \centering
    \includegraphics[width=\linewidth]{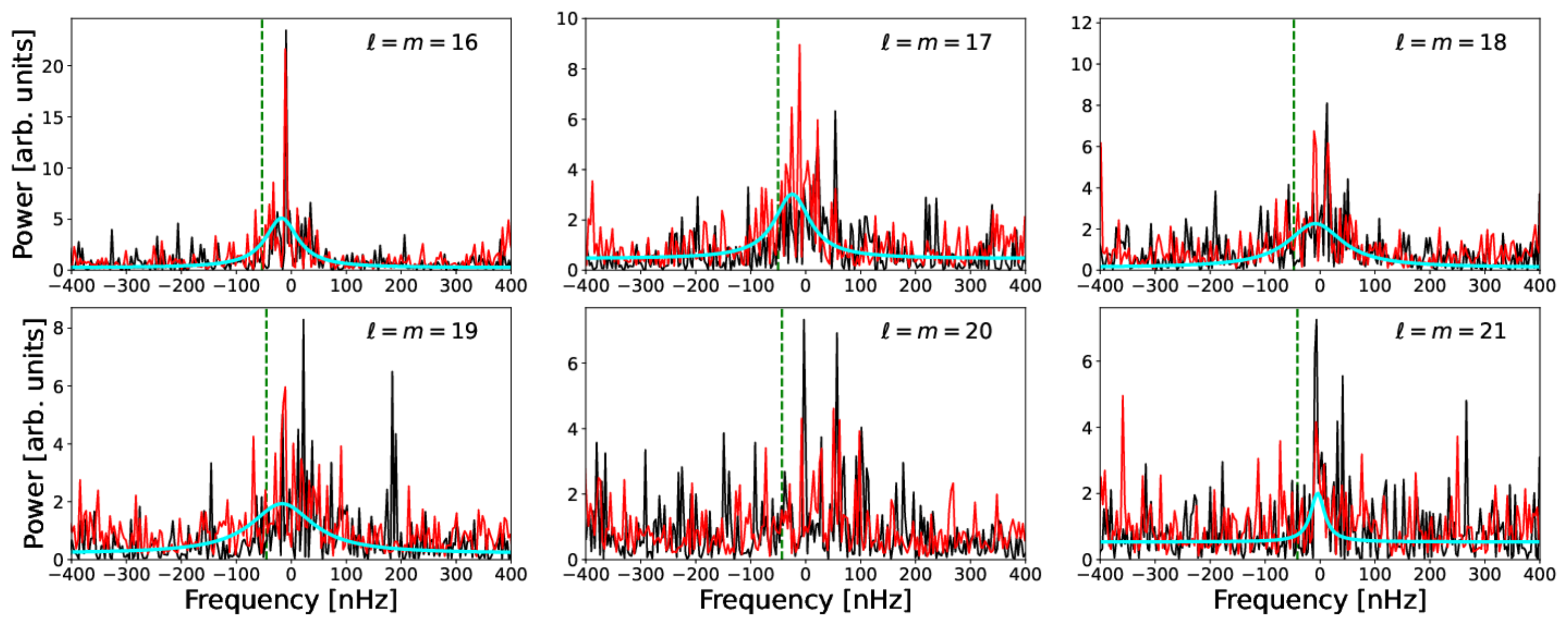}
    \caption{Comparison of the HMI MCA (black) and HMI RDA (red) radial vorticity power spectra for the same 10-year observation period (2010-2020). The harmonic degree $\ell$ and azimuthal order $m$ are shown in the top right of each panel. Lorentzian-like ridges are characterized as sectoral Rossby-Haurwitz waves. The expected mode frequency in the thin-shell approximation ($\omega=-2\Omega_0/(m+1)$) is shown by the green dashed lines. Lorentzian fits to the HMI RDA are shown by the cyan line.}
    \label{fig:spectra_slice}
\end{figure}

\begin{figure}
    \centering
    \includegraphics[width=0.7\linewidth]{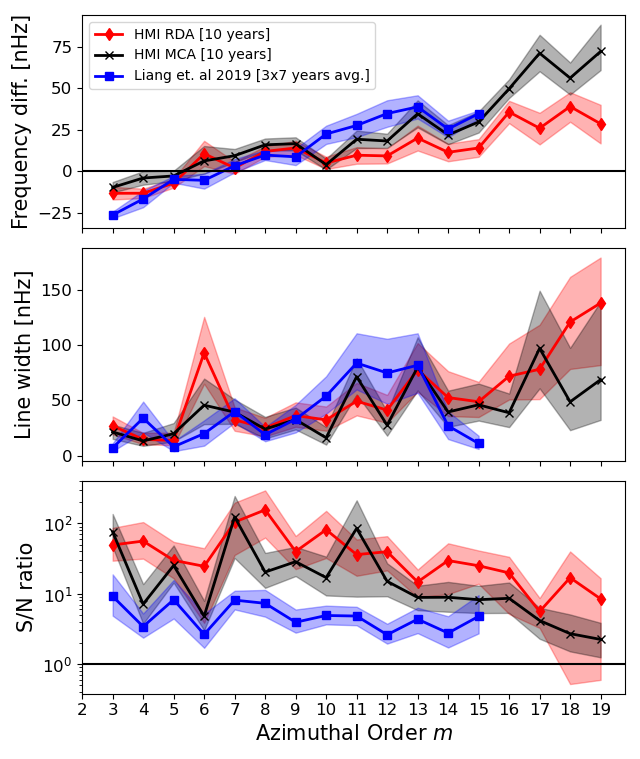}
    \caption{Plot of the mode fit parameters shown in Table~\ref{tab:modeparams}. We show the fits for the HMI RDA (red), HMI MCA (black) and for reference the measurements from time-distance power spectra \citep[][blue]{Liang2019}. Top panel: Difference between the measured mode frequency and the thin-shell dispersion relation. Middle panel: Measured line widths. Bottom panel: signal-to-noise ratio of the modes measured from the spectra. The shaded area shows the 68\% confidence level. }
    \label{fig:mode_params_plots}
\end{figure}

Using a Markov Chain Monte-Carlo method for fitting \citep[eg.][]{Hanson2022}, we fit the mode ridges for both HMI MCA and RDA and compute the associated errors. Figure~\ref{fig:spectra_slice} shows slices of the power spectra, and the fitted Lorenztians for $15< m< 22$. For $\ell=m=20$ the fitting routine failed to provide reliable results that were not dependent on the initial guess. This is due to low signal-to-noise ratio, and we do not provide the fit information. Table~\ref{tab:modeparams} shows the fit parameters and their respective $68$\% confidence intervals. Figure~\ref{fig:mode_params_plots} visualizes the mode fit parameters from Table~\ref{tab:modeparams}, as well as the modes $m>2$ which have been previously characterized. The mode frequencies from the RDA and MCA appear to have a trending difference beyond $m=10$. This results in the MCA modes appearing to travel prograde relative to the tracking rate. Systematic differences in the mode frequencies between different methods are well known, and we show the travel time measurements \citep{Liang2019} to further demonstrate this \citep[see also Fig.~3 of ][]{Hanson2020}. Small differences in the tracking rate, or low frequency noise, could lead to strong biases at large $m$. Further investigation is required to identify the differences between different data sets and methods, that lead to these discrepancies at high $m$. While there are differences in the mode frequencies, the line width measurements agree within errors. Finally, in the bottom panel of Fig.~\ref{fig:mode_params_plots} we show how the signal-to-noise ratio of the modes rapidly declines with increasing $m$, reaching values close to 1 beyond $m=19$.

\begin{deluxetable}{ccccc}
\label{tab:modeparams}
\tablecaption{Mode fit parameters of the Rossby-Haurwitz Sectoral modes measured from the HMI MCA and RDA radial vorticity power spectrum.}
\tablehead{$\ell=m$ & $\omega/2\pi$ [nHz] & Line Width [nHz] & signal-to-noise }
\startdata
 &  & \textbf{HMI RDA} &  & \\
16 & $-17.68^{6.75}_{6.72}$ & $72.08^{29.60}_{21.17}$ & $19.73^{13.67}_{14.59}$ \\ 
17 & $-24.47^{9.28}_{9.76}$ & $78.38^{40.35}_{27.14}$ & $5.58^{3.08}_{2.32}$ \\ 
18 & $-8.75^{8.62}_{9.18}$ & $121.25^{40.94}_{42.63}$ & $16.87^{22.95}_{16.35}$ \\ 
19 & $-16.78^{11.43}_{11.83}$ & $138.47^{41.27}_{56.33}$ & $8.43^{8.14}_{7.83}$ \\ 
20&-&-&-\\ 
21 & $-4.32^{8.88}_{7.01}$ & $30.83^{61.25}_{18.05}$ & $2.76^{3.25}_{1.40}$ \\ 

\\\\
& & \textbf{HMI MCA} & & \\
16 & $-3.64^{5.70}_{5.54}$ & $38.86^{17.45}_{13.04}$ & $8.58^{5.81}_{3.27}$ \\ 
17 & $20.75^{11.09}_{10.90}$ & $97.31^{52.35}_{36.27}$ & $4.14^{2.14}_{1.87}$ \\ 
18 & $8.42^{9.36}_{9.98}$ & $48.79^{48.94}_{25.73}$ & $2.69^{2.36}_{1.19}$ \\ 
19 & $26.83^{16.14}_{11.23}$ & $68.93^{70.87}_{36.57}$ & $2.23^{1.61}_{0.99}$ \\ 
20&-&-&-\\ 
21 & $2.32^{10.89}_{6.66}$ & $29.81^{31.01}_{17.25}$ & $3.46^{4.12}_{1.67}$ \\ 

\enddata
\tablecomments{Fits for each $m$ are performed in a frequency window $400$~nHz wide, centered on the dispersion relation $-2\Omega_0/(m+1)$. Negative frequencies indicate retrograde motion. We were unsuccessful in performing fits to the $\ell=m=20$ mode.}
\end{deluxetable}

\section{Results}

\subsection{Tracking at the equatorial rate}

\begin{figure}
    \centering
    \includegraphics[scale=0.8]{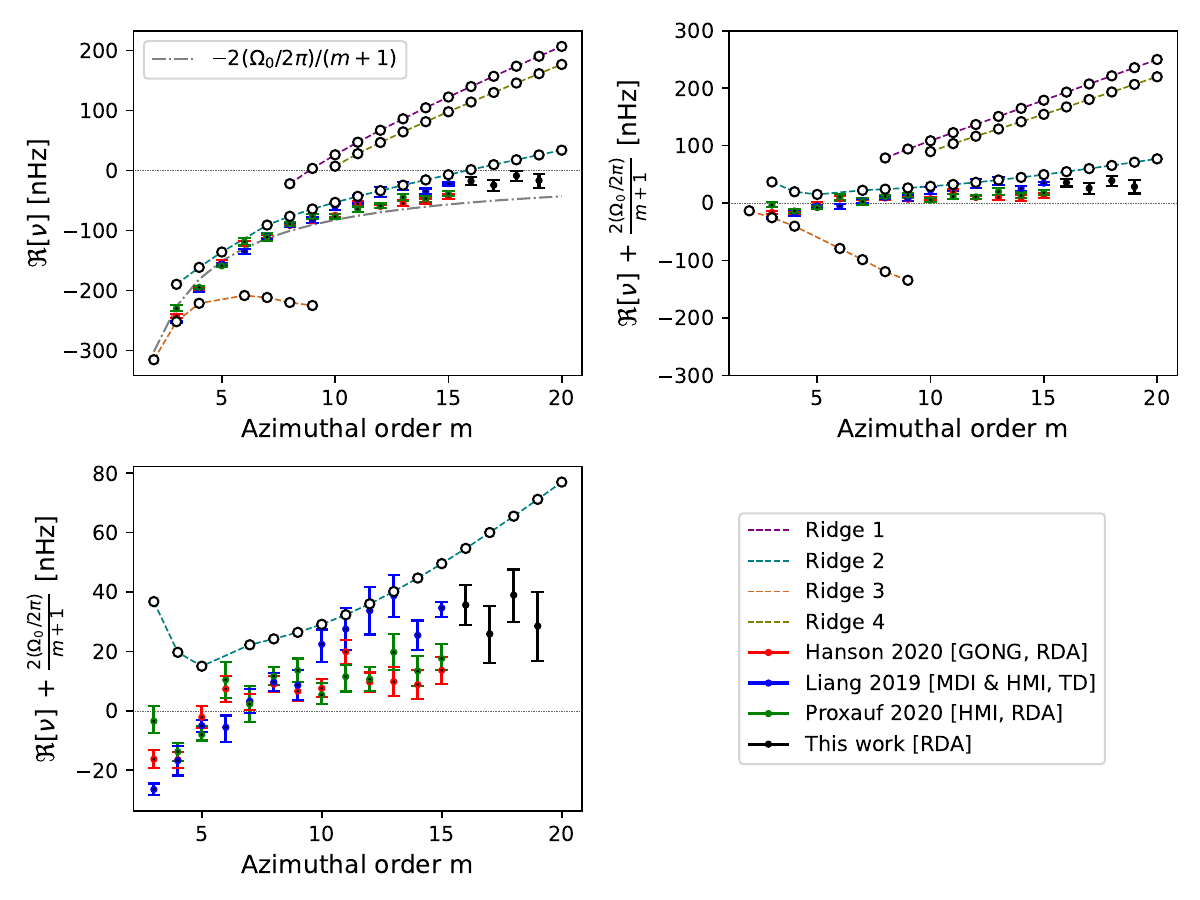}
    \caption{Top left panel: Spectrum of latitudinally symmetric solar inertial modes, obtained numerically (this work, white circles) compared with those from solar measurements. The error bars represent measured mode frequencies on the Sun reported by \citet{Liang2019,Proxauf2020,Hanson2020}, as well as new measurements in this work (black). Top right panel: Same as the left panel, with the frequencies corresponding to the Rossby-Haurwitz dispersion relation for sectoral modes $(\omega=-2\Omega_0/(m+1))$ subtracted from the eigenvalues. Bottom left panel: Zoom into the top right panel, focussing on the observed modes. In each panel, only a small section of modes that are the least-damped and have smooth eigenfunctions have been plotted.}
    \label{fig:spectrum}
\end{figure}

We have computed the spectrum of inertial modes by choosing $60$ Chebyshev polynomials along the radial direction, and $40$ harmonic degrees in the latitudinal direction. Since we can separate out symmetric and antisymmetric modes, the effective angular resolution doubles, and we only retain harmonic degrees corresponding to even $\ell+m$ while solving for the symmetric modes. We plot the spectrum of latitudinally symmetric (about the equator) modes in Figure \ref{fig:spectrum}. We only select sufficiently smooth solutions that have $90\%$ of their power concentrated below the radial Chebyshev order $n=10$ and the harmonic degree $\ell=m+15$ for a specific azimuthal order $m$. We have also restricted ourselves to modes that have two radial nodes at most. Alongside the spectrum that we obtain numerically, we have indicated the observed mode frequencies as measured by \citet{Liang2019} (time-distance (TD) analysis applied to combined HMI and MDI data, blue error bars), \citet{Hanson2020} (ring-diagram (RD) analysis applied to GONG data, red error bars), \citet{Proxauf2020} (ring-diagram analysis applied to HMI data, green error bars) and the new mode frequencies obtained by applying mode-coupling analysis to HMI data ((HMI, MC), denoted by black error bars). On the right panel, we plot the spectrum with the Rossby-Haurwitz dispersion relation $-2\Omega_0/(m+1)$ subtracted from the frequencies. Distinct ridges in the spectrum stand out, along which the eigenfunctions bear qualitative resemblance. We have indicated four of these ridges to which we refer as ``ridge-1'', ``ridge-$2$'',  ``ridge-$3$'' and ``ridge-$4$''. Ridge-$2$ lies strikingly close to the observed mode frequencies. 

\begin{figure}
    \centering
    \includegraphics[scale=0.5]{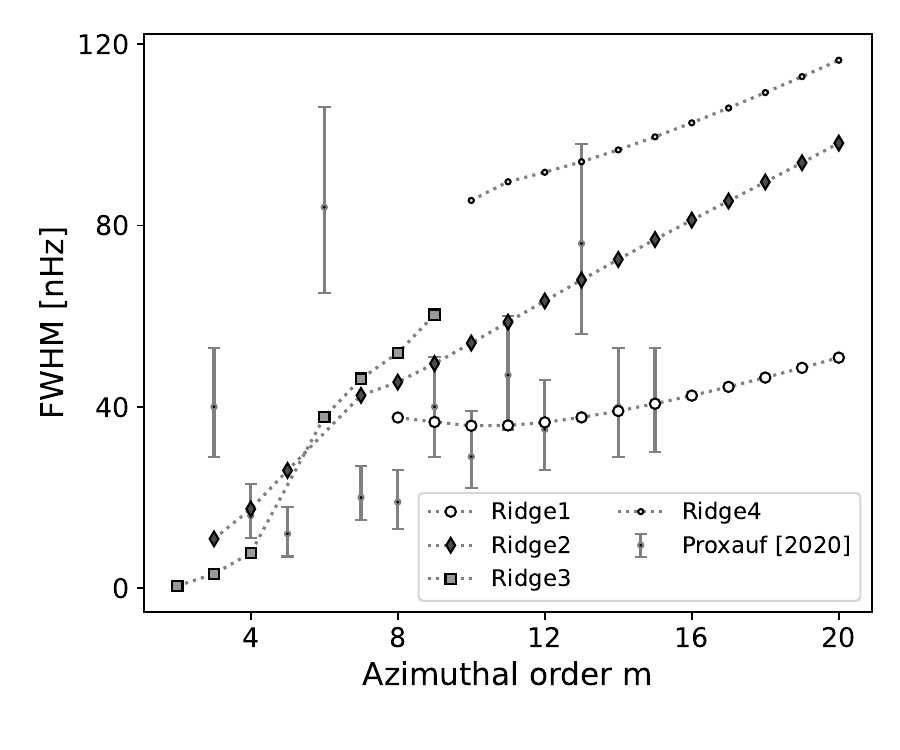}
    \caption{Full-width at half maximum of observed Rossby modes on the Sun (\citet{Proxauf2020}, error bars), and the doubled imaginary components of the numerically obtained eigenfrequencies for various ridges in Figure \ref{fig:spectrum}.}
    \label{fig:linewidth}
\end{figure}

We plot the doubled imaginary parts of the Rossby-ridge eigenfrequencies in Figure \ref{fig:linewidth}, along with the measured mode line-widths on the Sun. In this analysis, we have set the kinematic viscosity coefficient $\nu$ to $50\,\mathrm{km}^2/s$. We find that there is a correspondence between the observed line widths and the numerically obtained values, even though the level of turbulent damping chosen is less than the $100\,\mathrm{km}^2/s$ estimate by \citet{Gizon2021} for surface modes.

\begin{figure*}
    \centering
    \includegraphics[scale=0.4]{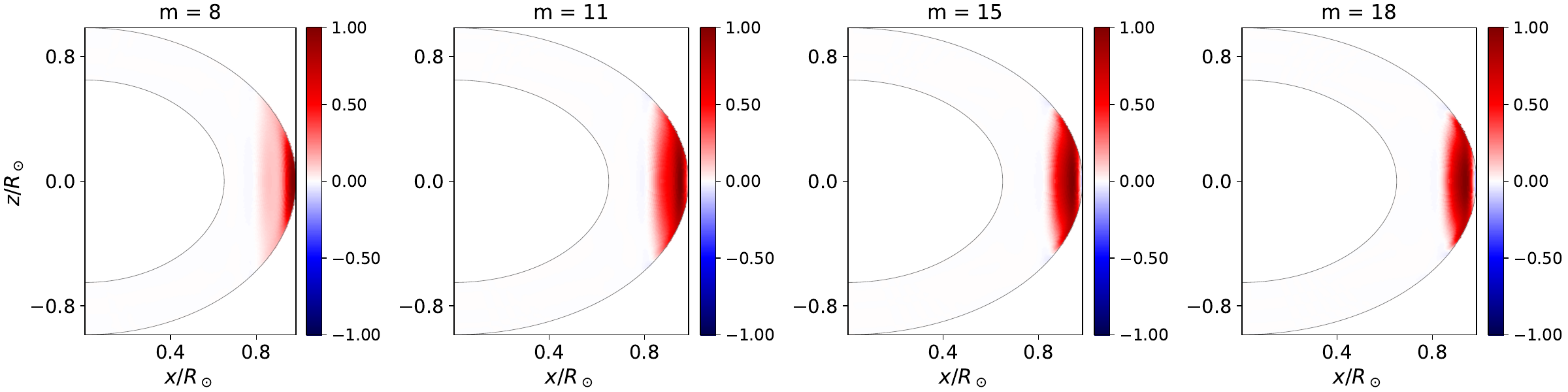}
    \caption{Real parts of the normalized toroidal stream function $V$ from "ridge $1$" for various values of $m$.}
    \label{fig:eig_diffm_ridge1}
\end{figure*}

\begin{figure*}
    \centering
    \includegraphics[scale=0.4]{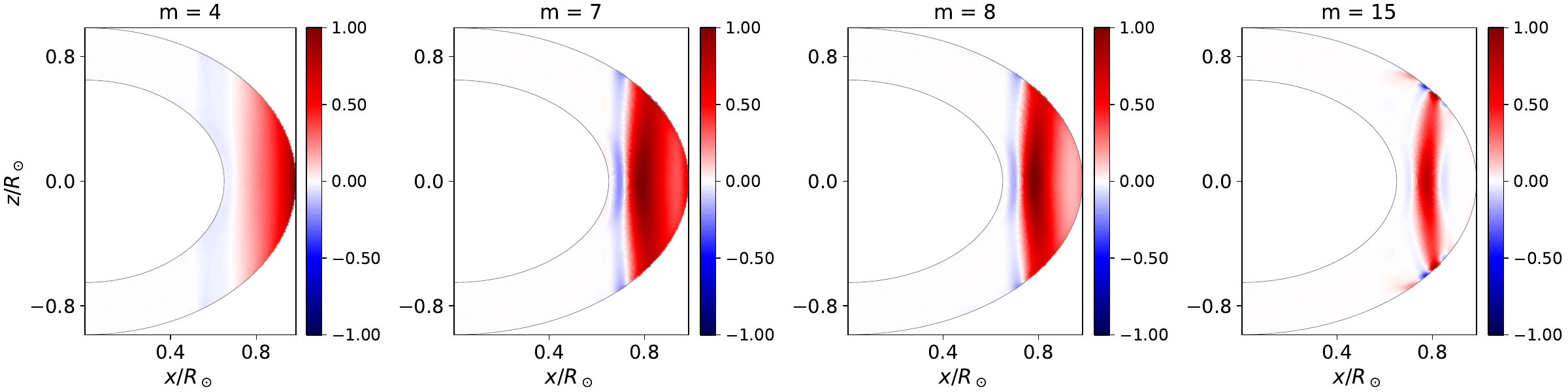}
    \caption{Real parts of the normalized toroidal stream function $V$ from "ridge $2$" for various values of $m$}
    \label{fig:eig_diffm_ridge2}
\end{figure*}

\begin{figure*}
    \centering
    \includegraphics[scale=0.4]{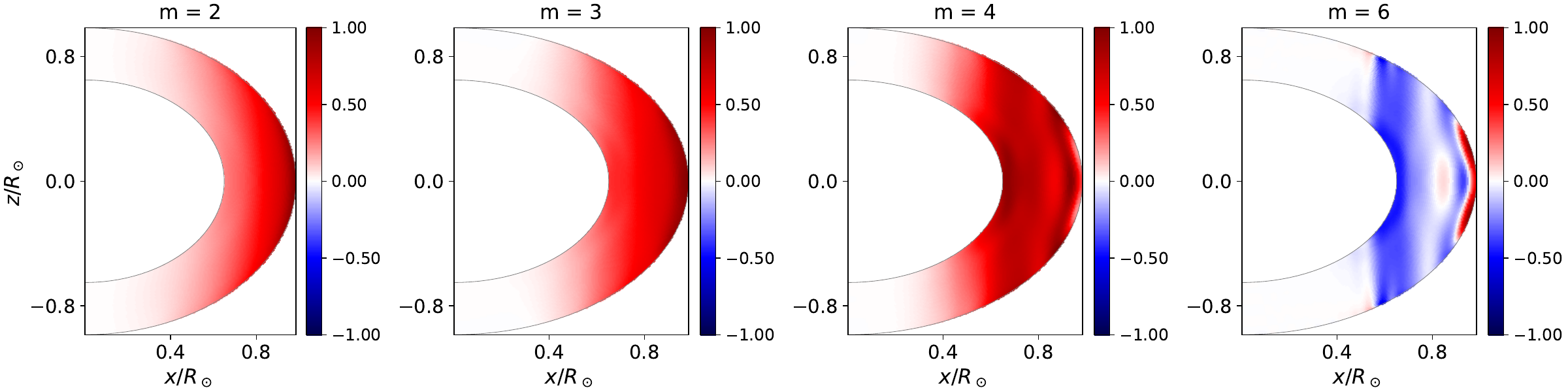}
    \caption{Real parts of the normalized toroidal stream function $V$ from "ridge $3$" for various values of $m$}
    \label{fig:eig_diffm_ridge3}
\end{figure*}

\begin{figure*}
    \centering
    \includegraphics[scale=0.4]{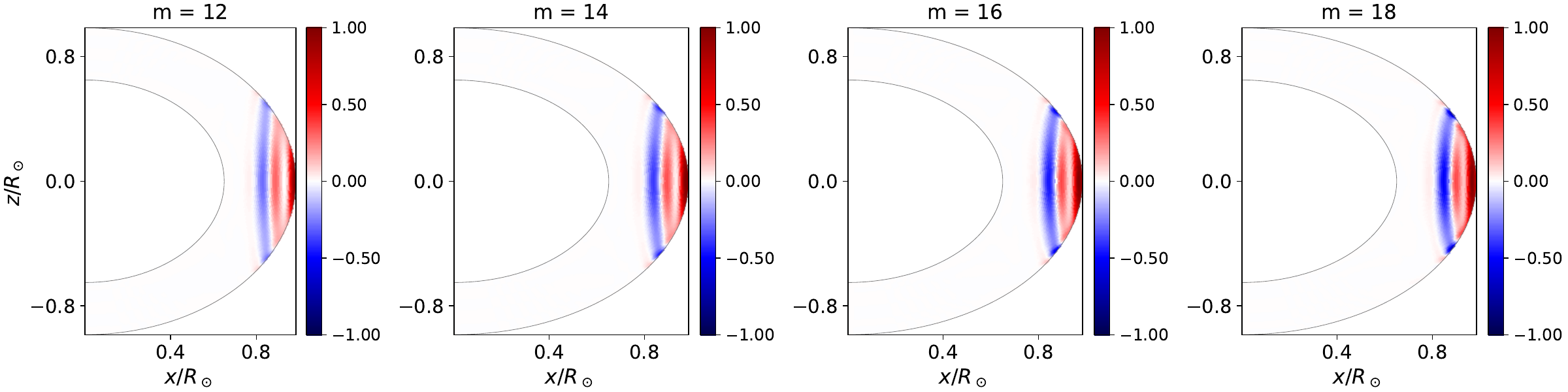}
    \caption{Real parts of the normalized toroidal stream function $V$ from "ridge $4$" for various values of $m$}
    \label{fig:eig_diffm_ridge4}
\end{figure*}

A subtlety that arises in interpreting the components of the eigenfunctions is that they are only determined to within an overall phase factor $\exp(i\psi)$, where $\psi$ is an arbitrary constant. To account for this, we ensure that the normalization process sets the $\left(n=0,\;\ell=m\right)$ component for the toroidal component $V$ to be purely real.
The eigenfunctions corresponding to ridge-$2$ appear to change their nature with increasing $m$, starting off as predominantly localized at the surface for $m\leq 4$ to peaking near the base of the convection zone for $m\geq 8$, with a transition around $m=6$. We plot the real part of the eigenfunction $V$ for various values of $m$ in Figure \ref{fig:eig_diffm_ridge2}. We show eigenfunctions for ridge $1$ in Fig. \ref{fig:eig_diffm_ridge1}, those for ridge-$3$ in Fig. \ref{fig:eig_diffm_ridge3} and for ridge-$4$ in Figure \ref{fig:eig_diffm_ridge4}. Ridge-$3$ eigenfunctions appear to change their nature with $m$ in a manner that is similar to those for ridge-$2$, with the functions being peaked at the surface at $m<4$, and progressively shifting to the bottom of the convection zone at higher $m$. However, this effect is much less pronounced than that for ridge-$2$, and at $m>4$, there appear to be numerical artifacts that develop at the surface, possibly originating from the lower boundary condition. The functions corresponding to ridge-$1$ as well as ridge-$4$ are of interest, as these are localized at the surface. In particular, the functions corresponding to ridge-$4$ display latitudinal nodes in a manner similar to the fitting function used by \citet{Proxauf2020}. However, the mismatch between the dispersion relation for ridge-$4$ and the data is considerable, so we refrain from establishing a correspondence between the two. At low $m$, eigenfunctions for ridge-$3$ appear to correspond to $n=0$, whereas those for ridge-$2$ correspond to the radial order $n=1$, although such an identification becomes difficult at higher $m$.

\begin{figure}
    \centering
    \includegraphics[scale=0.5]{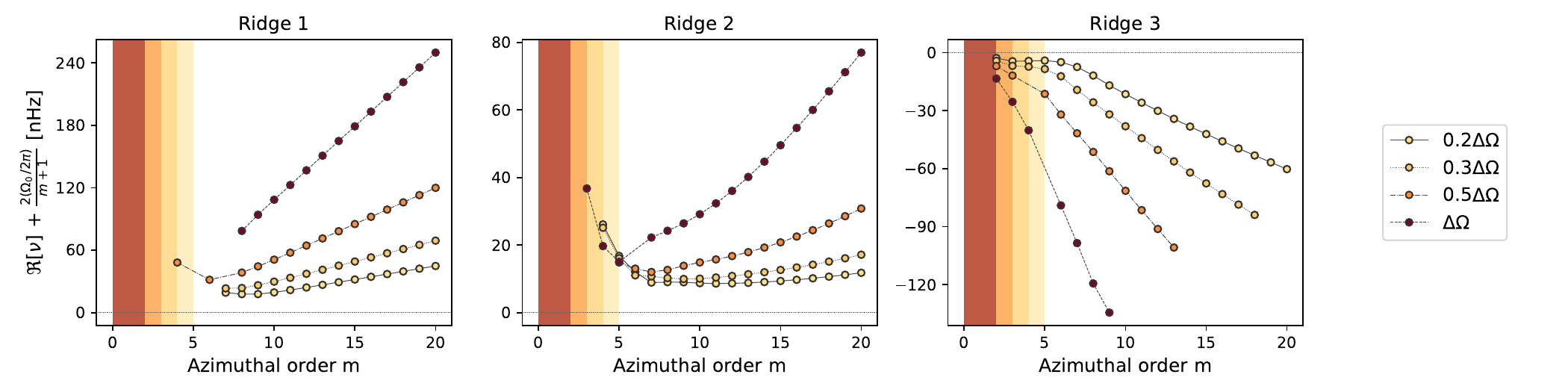}
    \caption{Spectra of symmetric equatorial Rossby modes obtained by scaling the differential rotation profile of the Sun by a constant factor. The panels correspond to different ridges, and the dots joined by lines in each subplot correspond to different rotation rates. The filled patches indicate the range in $m$ beyond which the sectoral Rossby modes are strongly impacted by differential rotation.}
    \label{fig:spec_fracvel}
\end{figure}

To trace the spectral ridges back to those for a uniformly rotating medium, we compute the inertial-mode spectrum by scaling the differential rotation $\Delta\tilde{\Omega}(r,\theta)$ by a constant $f$ for various values of $f$ lying in $(0,1)$. We plot the spectrum for four different values of $f$ in Figure \ref{fig:spec_fracvel}, after adding $2\Omega_0/(m+1)$ to the mode frequencies. We have highlighted three distinct ridges that progressively drift away from the Rossby-Haurwitz dispersion relation (indicated by the horizontal dotted line at an ordinate of zero). The largest deviation from the linear behavior is seen at low $m$, where the expectation that the $1/(m+1)$ dependence of the dispersion relation dominates the Doppler shift (which is proportional to $m$), and more so if the differential rotation rate is small compared to the reference rotation rate $\Omega_0$. The filled patches line in each sub-plot represents the lowest $m$ for which there exist critical latitudes, where the phase speed of sectoral Rossby modes equals the local differential rotation speed. Modes to the right of the filled patch line are influenced strongly by differential rotation, whereas the ones to the left are impacted weakly. We find that this roughly matches the point where ridge-$3$ starts to depart from the zero ordinate. For the solar rotation profile, critical latitudes exist for all $m>2$ \citep{Bekki2022b}, which explains the shape of ridge-3 in Figure \ref{fig:spectrum}. Despite this temptingly simple interpretation in terms of Doppler shifts, we note that not all the ridges appear to scale linearly with the factor $f$, and the eigenfunctions at high $m$ are significantly altered from those in a uniformly rotating medium.

We note that there is no reason to expect distinct ridges of eigenvalues that vary linearly with $m$, given that the eigenfunctions span a large spatial domain and sense a wide range of rotation rates. An analysis of the dispersion relations starting from the sensitivity kernels for these inertial modes may shed light on why this happens to be the case.

\subsection{Tracking at the Carrington rate}

\begin{figure}
    \centering
    \includegraphics[scale=0.6]{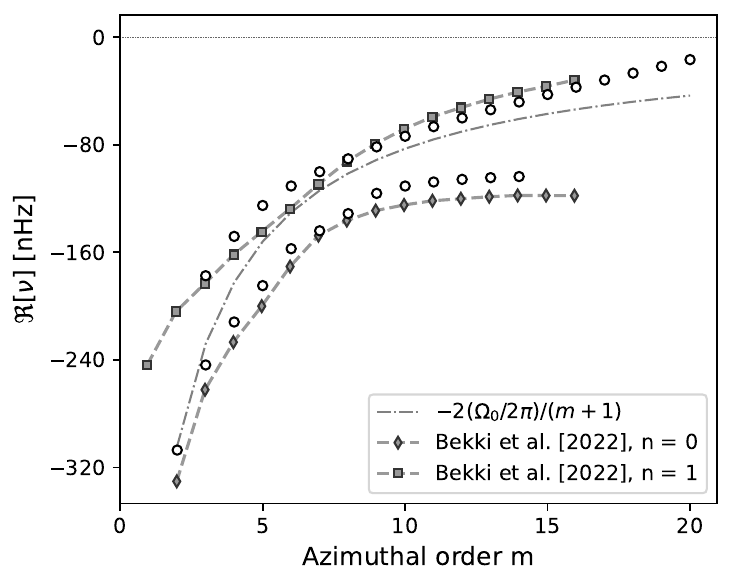}
    \caption{Equatorial inertial-mode spectrum on the Sun by setting the tracking rate to the Carrington rate (white circles). The diamonds and squares connected by dashed lines represent the dispersion relations from \citet{Bekki2022b}.}
    \label{fig:spectrum_bekki}
\end{figure}

We repeat the analysis in a truncated radial domain spanning $r=[0.71R_\odot,0.985R_\odot]$ to compare our results with \citet{Bekki2022a}. In this analysis, we truncate the solar rotation profile to the domain without scaling the radius, so the near-surface shear layer as well as the tachocline are both left out of the profile (although we find that including the near-surface shear layer does not impact the spectrum significantly). We further choose the tracking rate $\Omega_0$ to be the Carrington rotation rate $2\pi\times456$ nHz. We plot the spectrum in Figure \ref{fig:spectrum_bekki}. There is a correspondence between the ridges for $n=0$ and $n=1$ that are reported by \citet{Bekki2022b} and that in our result, although the match is not exact. Further work is necessary to improve the quantitative match between the two approaches by identifying the sources of systematic differences.

\section{Discussion and conclusion}

In this work, we have computed the spectrum of equatorial inertial modes in the Sun in the anelastic approximation, assuming a solar-like profile of differential rotation. We find that the mode frequencies often lie along distinct ridges. By tuning the strength of differential rotation, we may trace them back to those obtained in a uniformly rotating system, for which the dispersion relations are easier to interpret. We find that the spectral ridges vary nearly linearly with $m$, which suggests that the shifts in dispersion relations may be interpreted as those due to a constant effective rotation rate. This result is surprising because the spatial extent of the mode eigenfunctions would indicate that each mode ought to sample the rotation rate differently. Further study into the sensitivities of the modes to the rotation profile might shed light on this.

An interesting result that we obtain is that the sectoral ridge with the radial order $n=0$ --- which coincides with the canonical Rossby-Haurwitz dispersion relation $\omega=-2\Omega_0/(m+1)$ for a uniformly rotating medium --- is strongly deflected in the retrograde direction. It is likely, therefore, that the observed mode frequencies correspond to one of the other ridges that we identify in the spectrum. A similar conclusion was reached by \citet{Bekki2022a}, where the authors suggested that the observed frequencies might correspond to modes with $n=1$. Based on the proximity of the mode frequencies to the measured frequencies, we speculate this to be the case as well. However, we note that the eigenfunctions for these modes are localized near the base of the convection zone for $m>8$, unlike the near-surface measurements on the Sun. Moreover, the mode frequencies at $m<7$ for ridge-$2$ appear to differ considerably from solar measurements, although this might be because these modes extend deep into the Sun and are therefore more sensitive to the lower boundary condition. Ridges $1$ and $4$, on the other hand, have eigenfunctions that are localized near the surface, and, in particular, the latter has eigenfunctions that seem to resemble the latitudinal form suggested by \citet{Proxauf2020}. However, the dispersion relation for modes along this ridge lies considerably away from the observed frequencies, so these are unlikely to be candidates for the observed modes.

To get a complete picture of inertial modes under the effect of differential rotation, we have also presented the measurements of solar sectoral Rossby-Haurwitz waves with azimuthal orders greater than 15, using ring-diagram and mode-coupling analysis. These modes have eluded previous studies due to signal-to-noise or aliasing effects, and the first measurements to our knowledge were reported recently by \citet{Waidele2023} using time-distance seismology. The most interesting finding from these measurements is the increasing discrepancy between the thin-shell dispersion relation and the observations. While MCA and RDA have a small disagreement --- which warrants further investigation into data sets and methodology --- they both suggest a strong divergence of the solar modes from the theory towards less-retrograde frequencies. This trend with increasing $m$ is qualitatively similar to the behavior of the ridge-2 from our models.

A few words are in order on the questions left untouched in this work. We have focussed strictly on the equatorial modes, ignoring the wider spectrum of critical-latitude and high-latitude modes that have purportedly been detected. We have also chosen to focus on modes with $m>1$, which excludes the high-latitude spiral mode that was detected by \citet{Hathaway2021}. Moreover, we have limited our analysis strictly to equatorially symmetric modes, thereby omitting the high-frequency Rossby modes that were detected by \citet{Hanson2022}. Further analysis using this approach may choose to focus on these aspects. Additionally, our analysis is purely hydrodynamic, but magnetic fields may play an important role governing the dispersion relation \citep{Dikpati2020}. In particular, total vorticity is no longer conserved in the presence of Lorentz forces \citep{Zaqarashvili2021}, although the impact of this may not be significant in the solar near-surface layers, where the dispersion relation of the observed modes appear to closely resemble that of classical Rossby waves. The approach presented here may be extended to include magnetic fields, which might unveil a richer spectrum of modes, such as atmospheric modes \citep{McIntosh2017}, as well as those that may exist in the solar tachocline \citep{Zaqarashvili2010,Zaqarashvili2018,Dikpati2018}. 

\acknowledgments
This material is based upon work supported by Tamkeen under the NYU Abu Dhabi Research Institute grant G1502. We also acknowledge support from
the King Abdullah University of Science and Technology (KAUST) Office of Sponsored Research (OSR) under award OSR-CRG2020-4342.
This research was carried out on the High-Performance Computing resources at New York University Abu Dhabi.

\appendix
{}
\section{Computing matrix elements\label{app:matrixelem}}

We consider a space $S$ spanned by the normalized associated Legendre polynomials $\hat{P}_{\ell m}\left(\cos\theta\right)$ having a specific azimuthal order $m$, where we omit the Condon-Shortley phase factor in the definition.
We denote the matrix elements of an operator $O\left(\theta\right):S\rightarrow S$ in a basis of normalized associated Legendre polynomials $\hat{P}_{\ell m}\left(\cos\theta\right)$ as
\begin{align}
    O_{\ell \ell^\prime, m} = \int_{-1}^{1} d\left(\cos\theta\right)\, \hat{P}_{\ell m}\left(\cos\theta\right) O\left(\theta\right) \hat{P}_{\ell^\prime m}\left(\cos\theta\right).
\end{align}
In general, we compute the matrix element of such an operator by exploring a correspondence with Jacobi polynomials, although in the special case where $O(\theta)$ is a function, the matrix element may also be evaluated in terms of Wigner-$3j$ symbols by utilizing the $3$-term integral relations for spherical harmonics.

We start by recognizing that normalized associated Legendre polynomials $\hat{P}_{\ell m}\left(x\right)$ are identical to the weighted and normalized ultraspherical (Gegenbauer) polynomials $(1-x^2)^{m/2}\,\hat{C}^{m+1/2}_{\ell-m}\left(x\right)$. One way to understand this correspondence is that the Legendre polynomial $P_\ell(x)$ is equal to the ultraspherical polynomial $C^{1/2}_{\ell}(x)$. The $m$-th derivative of $C^{1/2}_{\ell}(x)$ is equal to $C^{m+1/2}_{\ell-m}(x)$ within a normalization factor. Since the associated Legendre polynomial $P_{\ell m}(x)$ is proportional to the $m$-th derivative of the Legendre polynomials $P_\ell(x)$ weighted by $(1-x^2)^{m/2}$, the correspondence between the former and the weighted ultraspherical polynomials is naturally established. The matrix element $O_{\ell\ell^\prime,m}$ may therefore be obtained by evaluating the corresponding matrix element in a basis of the weighted ultraspherical polynomials instead. The latter permits fast orthogonal polynomial transforms \citep{Olver2013}, and an algorithm to obtain the coefficients in the weighted ultraspherical basis is already implemented in the Julia package ApproxFun.jl. We borrow the results directly in our application.

\bibliographystyle{aasjournal}
\bibliography{references}

\end{document}